\documentclass[amsmath,superscriptaddress,reprint,aps]{revtex4-2}
\usepackage{graphicx}
\usepackage{dcolumn}
\usepackage{bm}
\usepackage{hyperref}
\usepackage{siunitx}
\usepackage{color, soul}

\setstcolor{red}

\begin{document}
	
	
	\title{Record-high-Q AMTIR-1 microresonators for mid- to long-wave infrared nonlinear photonics}

     \author{Liu~Yang}
      	\affiliation{Department of Physics, Faculty of Science and Technology, Keio University, Yokohama, 223-8522, Japan}
	\affiliation{Department of Electronics and Electrical Engineering, Faculty of Science and Technology, Keio University, Yokohama, 223-8522, Japan}
    \affiliation{Electronic Materials Research Laboratory, Key Laboratory of the Ministry of Education $\&$ International Center for Dielectric Research, School of Electronic Science and Engineering, Faculty of Electronic and Information Engineering, Xi’an Jiaotong University, Xi’an, 710049, China}
    
              \author{Ryo~Sugano}
	\affiliation{Department of Electronics and Electrical Engineering, Faculty of Science and Technology, Keio University, Yokohama, 223-8522, Japan}
    
              \author{Ryomei~Takabayashi}
	\affiliation{Department of Electronics and Electrical Engineering, Faculty of Science and Technology, Keio University, Yokohama, 223-8522, Japan}

         \author{Hidetoshi~Kanzawa}
	\affiliation{Department of Physics, Faculty of Science and Technology, Keio University, Yokohama, 223-8522, Japan}

     \author{Hajime~Kumazaki}
	\affiliation{Department of Physics, Faculty of Science and Technology, Keio University, Yokohama, 223-8522, Japan}
    \affiliation{Department of Electronics and Electrical Engineering, Faculty of Science and Technology, Keio University, Yokohama, 223-8522, Japan}
    
            \author{Yongyong~Zhuang}
    \affiliation{Electronic Materials Research Laboratory, Key Laboratory of the Ministry of Education $\&$ International Center for Dielectric Research, School of Electronic Science and Engineering, Faculty of Electronic and Information Engineering, Xi’an Jiaotong University, Xi’an, 710049, China}

                \author{Xiaoyong~Wei}
    \affiliation{Electronic Materials Research Laboratory, Key Laboratory of the Ministry of Education $\&$ International Center for Dielectric Research, School of Electronic Science and Engineering, Faculty of Electronic and Information Engineering, Xi’an Jiaotong University, Xi’an, 710049, China}

    \author{Takasumi~Tanabe}
	\affiliation{Department of Electronics and Electrical Engineering, Faculty of Science and Technology, Keio University, Yokohama, 223-8522, Japan}

	\author{Shun~Fujii}
	\email[Corresponding author. ]{shun.fujii@phys.keio.ac.jp}
 	\affiliation{Department of Physics, Faculty of Science and Technology, Keio University, Yokohama, 223-8522, Japan}
	
	
\begin{abstract}
 AMTIR-1 chalcogenide glass has shown its potential for use in thermal imaging systems owing to its low refractive index, thermal resistance and high transparency across the infrared wavelength regime.  Here we report a millimeter-scale high-Q whispering gallery mode microresonator made of AMTIR-1. The recorded Q-factor has reached $1.2\times10^7$ at 1550~nm, which is almost two-orders of magnitude higher than previously reported values. We characterize the thermal properties, where low thermal conductivity plays an important role in thermal resonance tuning.  We further show that AMTIR-1 resonators support anomalous dispersion as well as a low absorption coefficient near the 7~\textmu m wavelength band, thus offering the possibility of providing suitable platforms for mid-infrared, long-wave infrared nonlinear optics including microresonator frequency comb generation.
\end{abstract}

\maketitle

High-quality (Q) factor whispering gallery mode (WGM) microresonators have been utilized as attractive platforms for studying nonlinear and quantum optics since the high-Q properties and large nonlinear coefficients enable various nonlinear optical effects such as four-wave mixing (FWM), harmonic generation, and stimulated Raman scattering, at very low pump powers~\cite{Herr2012,Fujii2018,Lin2022}. In particular, optical parametric oscillation based on the FWM process allows the generation of quantum squeezing states and coherent frequency combs from a continuous-wave input, leading to intriguing practical applications such as secure quantum communications~\cite{Yang2021,Zhang2023}, and gas and molecular spectroscopy~\cite{Kovalchuk2001,Zhang2013} in miniature devices. 

Frequency comb generation in the infrared region is especially attractive as it permits access to the molecular absorption region, where there are vibrational absorption signatures for many chemical species. In fact, combs with a large line spacing and a very broad span are required for efficient vibrational spectroscopy~\cite{Ideguchi2013,Hashimoto2019}. Despite such a luminous landscape, few microresonator-based applications in the mid-infrared and long-wave infrared regimes have been developed due to the lack of high-Q resonator platforms that have both ultra-low loss properties and suitable dispersion profiles, both of which are essential for nonlinear processes. Although there have been several demonstrations of microresonator frequency comb generation in silicon nitride ($\mathrm{Si_3N_4}$)~\cite{Luke2015,Guo2018}, silicon (Si)~\cite{Griffith2015,Yu2018}, and fluoride crystals~\cite{Wang2013,Savchenkov2015,Wu2022} in the mid-infrared region, with these platforms it is difficult to extend the wavelength beyond 6~\textmu m because of strong material absorption and the limitations of dispersion engineering. This fact poses a challenge as regards direct access to the fingerprint region, i.e., 6.7-20 \textmu m ($1500-500~\mathrm{cm^{-1}}$), where each molecule exhibits a unique absorption pattern in contrast to the functional group region, i.e., 2.5-6.7~\textmu m.

AMTIR-1 ($\mathrm{Ge_{33}As_{12}Se_{55}}$), a chalcogenide glass, is a  material that is widely used for infrared optical components thanks to its various desirable properties in the infrared regime including a wide transmission window ($\sim$0.75--14 \textmu m), a low two-photon absorption coefficient, a low refractive index ($n\sim2.45-2.5$) and a high optical nonlinearity ($n_2=10^{-17}-10^{-18}~\mathrm{m^2/W}$)~\cite{Ensley2019,Choi2007,Singh2015}.  In particular, the nonlinear refractive index $n_2$ is two or three orders of magnitude higher than that of fluoride crystals and silica glass~\cite{Ensley2019}. The photosensitivity is also a unique property of this material, and has been exploited for a wide variety of applications~\cite{Singh2015,Lee2007,Lee2009,Lee2010}.
Although there have been several studies on AMTIR-1 microcavities over the past two decades ~\cite{Choi2007,Singh2015,Lee2007,Lee2009,Lee2010,Conteduca2015}, the high nonlinearity and low absorption coefficients remain largely unexplored for nonlinear photonic applications.
Meanwhile, the reported Q-factors are limited to moderate values ($\sim10^5$), which may have caused us to overlook the potential of this material.

In this work, we propose a monolithic AMTIR-1 WGM microresonator as a desirable platform for nonlinear applications in the mid-infrared and long-wave infrared regions. We report the experimental observation of a high Q value of $1.2\times10^7$ at 1550~nm, which is almost two orders of magnitude higher than that of previously reported microcavities made of AMTIR-1.  The thermal properties of AMTIR-1 resonators are determined experimentally, and the results are consistent with numerical simulations that take account of thermal conduction coefficients. We show that the dispersion profile of AMTIR-1 resonators allows broadband frequency comb generation above 7~\textmu m, thus revealing its great potential as a coherent light source in the molecular fingerprint region.

\begin{figure}[t!]
\centering\includegraphics{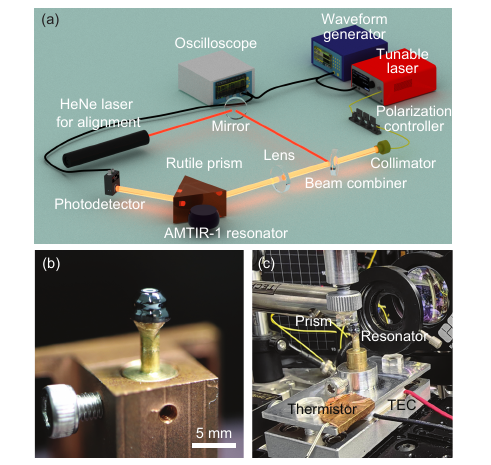}
\caption{\label{Fig_setup} (a) Schematic diagram of the experimental setup. (b) Photograph of an AMTIR-1 microresonator. (c) Sideview of the prism coupling system, where the resonator temperature is controlled with a thermoelectric cooler.}
\end{figure}

The experimental setup for a Q-factor measurement is shown schematically in Fig.~\ref{Fig_setup}(a). 
A microresonator is fabricated from a commercially available optical grade AMTIR-1 window with a thickness of 3~mm by manual shaping and surface polishing [Fig.~\ref{Fig_setup}(b)]. The details of the fabrication procedure is presented in Supplement~1. The resonator has a free-spectral range (FSR) of 9.7~GHz, with a diameter of $\sim$3.9~mm. A wavelength tunable, narrow-linewidth ($\sim$200~kHz) laser (Santec, TSL-570) operating in the C-band wavelength range is utilized to measure the transmission spectrum of the AMTIR-1 resonator, where a polarization controller is used to adjust the polarization of the input light. The resonator is mounted on a 3D translation stage with 20~nm resolution (Thorlabs, MAX312D/M) for precise coupling control. Given the refractive index of AMTIR-1 ($n\sim$2.545 at 1550~nm), a right-angle rutile prism (Thorlabs, ADT-6) is used to obtain efficient coupling as shown in Fig.~\ref{Fig_setup}(c). A HeNe laser is incorporated to provide a reference optical path for alignment. The transmitted light is collected by a photodetector (Thorlabs, PDA10CS2) and recorded with a digital sampling oscilloscope.

\begin{figure}[t!]
\centering\includegraphics{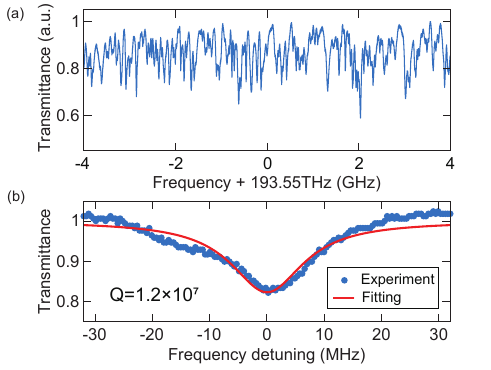}
\caption{\label{Fig_transmission} (a) Observed transmission spectrum and (b) the Lorentzian fitting, yielding a loaded Q-factor of up to $1.2\times10^7$. The measured mode belongs to the TE modes.}
\end{figure}

\begin{figure}[t!]
\centering\includegraphics{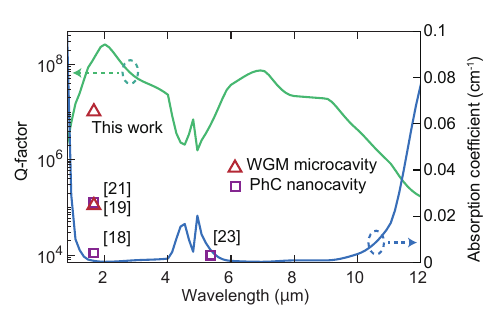}
\caption{\label{Fig_absorption} Reported Q-factors as a function of wavelength of microcavities made of AMTIR-1. WGM microcavity, ref.~\cite{Singh2015}, Photonic crystal (PhC) nanocavity, refs.~\cite{Choi2007,Lee2009,Conteduca2015}. The solid blue line denotes the absorption coefficient of AMTIR-1 (right axis), and the solid green line corresponds to the Q-factor calculated with the material absorption (left axis).}
\end{figure}

A series of WGMs are excited when evanescent light coupling is achieved by the careful adjustment of the coupling location. It should be noted that we only observed TE modes because the phase-matching condition can be met for the cavity modes and the extraordinary refractive index of the rutile prism ($n_e=2.691$). Figure~\ref{Fig_transmission}(a) shows a typical transmission spectrum. Higher-order transverse modes are observed within the 1-FSR. Among the many observed modes, we select one representative mode to evaluate the Q-factor (See also Supplement~1 for details). Lorentzian fitting of its spectrum yielded a full width at half maximum (FWHM) of 16.1~MHz, corresponding to a loaded $Q$ of $1.2\times10^7$ (Fig.~\ref{Fig_transmission}(b)). The frequency axis is calibrated with the reference signal of a fiber-based Mach-Zehnder interferometer~\cite{Fujii2023}. We note that the recorded Q, which  exceeds  10~million in the telecom band, is the highest value observed in microcavities made of AMTIR-1 among the previously reported values for a microdisk ($Q=1.2\times10^5$)~\cite{Singh2015} and a photonic crystal ($Q=1.25\times10^5$)~\cite{Lee2009}. Figure~\ref{Fig_absorption} summarizes the absorption-limited Q-factor and corresponding absorption coefficient with experimentally reported Q-factors of AMTIR-1 microcavities. According to the absorption coefficients over the infrared regime, the estimated absorption loss is $\alpha \approx 0.001~\mathrm{cm^{-1}}$ at 1550~nm, giving rise to an approximate absorption Q of $1\times10^8$, derived from a simple formula, $Q_\mathrm{abs} = 2\pi n/(\lambda \alpha)$.

AMTIR-1 exhibits its lowest absorption property in the specific infrared regime from 6~\textmu m to 8~\textmu m, and therefore we anticipate that the resonator will exhibit a high Q value of up to a maximum of $\sim8 \times 10^7$ in the wavelength window.

We next focus on the thermal properties of AMTIR-1 resonators. The resonance frequency of optical microresonators can be influenced by two primary thermal effects, namely the thermo-optic and thermal expansion effects. The former arises from the change in the refractive index, while the latter results from material expansion. Both effects simultaneously change the optical path length of the resonator, thus altering the resonance frequencies and the free-spectral range~\cite{Fujii2023}. Compared with other optical materials, AMTIR-1 exhibits a relatively high thermo-optic coefficient ($dn/dT = 78\times 10^{-6}~\mathrm{K^{-1}}$) and thermal expansion coefficient ($\beta = (1/l)dl/dT = 12.6\times 10^{-6}~\mathrm{K^{-1}}$) at 1550~nm, which allow highly efficient frequency tuning. 
The resonance frequency shift $\Delta f$ with a temperature increase $\Delta T$ is described by the following equation:
\begin{equation}
  df/dT =  1/n  (dn/dT) \Delta T +  \beta \Delta T, 
\end{equation}
which gives a frequency tuning efficiency  of $df/dT=$8.39~GHz/K at room temperature.

\begin{figure}[t!]
\centering\includegraphics{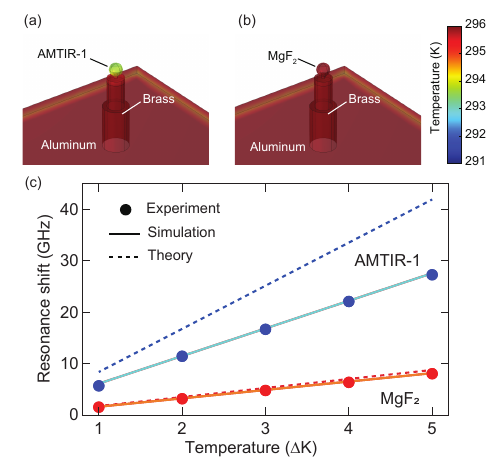}
\caption{\label{Fig_thermal} (a, b) Simulated steady-state temperature distributions when the temperature of the aluminum base is increased from 291~K to 296~K for an AMTIR-1 resonator and an $\mathrm{MgF_2}$ resonator, respectively. (c) Resonance frequency shifts as a function of temperature change. Experimental results (filled circles) align with simulated results (solid lines) rather than theoretical predictions (dashed lines), especially in an AMTIR-1 resonator due to its low thermal conductivity.}
\end{figure}

With this prediction in mind, we measure the thermal tuning efficiency by tracking resonance frequencies while changing the cavity temperature. In the experiment, a thermoelectric cooler is placed beneath the resonator to regulate its temperature [Fig.~\ref{Fig_setup}(c)], and the set temperature is gradually increased from 291~K to 296~K in 1~K steps to minimize the effect of cavity expansion on the coupling condition. The transmission spectrum is then recorded after reaching thermal equilibrium. It should be noted that the experiment is conducted in a darkroom to avoid the influence of ambient light since AMTIR-1 is a photosensitive material. Figure~\ref{Fig_thermal}(c) shows the experimental result, which yields an extracted tuning efficiency of 5.39~GHz/K in an AMTIR-1 resonator This is significantly smaller than the theoretical value of 8.39~GHz/K. 

To understand this large discrepancy, we perform a heat transfer simulation with the finite-element method, where we take thermal conduction into account. When the temperature is set at 296~K, that is 5 degrees higher than room temperature (291~K),  a resonator mounted on a brass jig and an aluminum base quickly reach the target temperature. However, the resonator only reaches approximately 294.5~K due to the low thermal conductivity of AMTIR-1 ($K=0.25~\mathrm{W/(m\cdot K)}$), which causes the difference between the experimental and theoretical values. Consequently, the simulation for each temperature step yields a thermal tuning efficiency of 5.37~GHz/K for AMTIR-1 resonators, and this value closely aligns with the experimental result. For comparison, we conduct the same experiment and simulation for an $\mathrm{MgF_2}$ resonator. The result indicates that the simulation closely reproduces the experiment, while the theoretical value that considers only the thermo-optic and thermal expansion coefficients also  agrees well with the experimental value owing to the large conductivity of up to $\sim$33.6~$\mathrm{W/(m\cdot K)}$ at room temperature~\cite{Fujii2023}.

Finally, we investigate the potential of high-Q AMTIR-1 resonators for nonlinear applications. By taking advantage of its low absorption and high nonlinear coefficients in the mid-infrared and long-wave infrared regimes, AMTIR-1 can be used as a resonator platform for frequency comb generation. As the threshold power for four-wave mixing scales to $\sim V/(n_2 Q^2)$, where $V$ is the mode volume, the high nonlinear index makes up for the Q factors being lower than that of ultrahigh-Q ($>10^8$) resonators~\cite{Herr2012}. A dissipative Kerr soliton (DKS) microcomb seeded by a continuous-wave laser can form a coherent, broad optical spectrum, which promises significant advantages for mid-IR sensing applications~\cite{Yu2018}. However, one critical requirement for soliton formation is a suitable anomalous dispersion that balances the Kerr nonlinearity to sustain the pulse waveform~\cite{Herr2014}. 

\begin{figure}[t!]
\centering\includegraphics{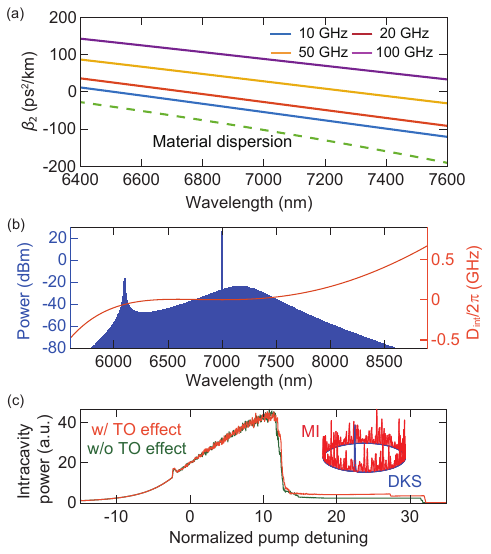}
\caption{\label{Fig_dispersion} (a) Simulated dispersion $\beta_2$ of AMTIR-1 resonators for different FSRs and material dispersion. The curvature radius is 50~\textmu m. (b) Simulated optical spectrum and integrated dispersion $D_\mathrm{int}$ with a center wavelength of  7~\textmu m. Other parameters are as follows: $Q_\mathrm{int}=1\times10^7$, $Q_\mathrm{ext}=3\times10^7$ , $D_1/2\pi=20$~GHz, $D_2/2\pi=12.3$~kHz, $D_3/2\pi=-185$~Hz, $g=1.7\times10^{-3}$, and $P_\mathrm{in}=|A_\mathrm{in}|^2=500$~mW.
(c) Evolution of intracavity power with and without the TO effect. The detuning is normalized by a half of the decay rate (i.e., $2\delta_0/\kappa_\mathrm{tot}$). The inset shows temporal waveforms of the MI (red) and DKS (blue) states.}
\end{figure}

Figure~\ref{Fig_dispersion}(a) shows the simulated group-velocity dispersion of AMTIR-1 resonators for different FSRs. Due to the strong material dispersion, AMTIR-1 resonators are in a normal dispersion regime at the near- and mid-infrared wavelengths.  We find that the geometric dispersion finally balances the material dispersion above 6~\textmu m, resulting in moderate anomalous dispersion profiles that are ideal for broadband comb generation around 7~\textmu m for a certain FSR.  We also note that AMTIR-1 exhibits its lowest absorption coefficient at this wavelength, which will guarantee high-Q properties even in such long-wave infrared regions (Fig.~\ref{Fig_absorption}).

Using the Lugiato-Lefever equation (LLE), we conduct numerical simulations of soliton frequency comb generation at a center wavelength of 7~\textmu m. The mean-field LLE is expressed as~\cite{Coen2013,Fujii2018}:
\begin{equation}
\begin{split}
\frac{\partial A(\phi,t)}{\partial t} = -\left( \frac{\kappa_{\mathrm{tot}}}{2} + i \delta_0 \right)A -i  \sum_{k=2} \frac{D_k}{k!} \left( \frac{\partial}{i \partial \phi} \right) ^k A \\
+ i g |A|^2 A + \sqrt{\kappa_{\mathrm{ext}}}A_{\mathrm{in}}, \label{Eq.1}
\end{split}
\end{equation}
where $t$ is the time describing the evolution of the field envelope, $\phi $ is the azimuthal angular coordinate. The total decay rate $\kappa_\mathrm{tot}=\omega_0/ Q$ is given by the sum of the intrinsic loss $\kappa_\mathrm{int}$ and the external coupling rate $\kappa_\mathrm{ext}$. $\delta_0$ and $g$ represent the pump detuning from the resonance, and a nonlinear coefficient, respectively.  The resonator is driven by a continuous-wave input, $A_{\mathrm{in}}=\sqrt{P_{\mathrm{in}}/\hbar \omega_p}$. The dispersion term $D_k$ includes higher-order dispersion but the second-order term $D_2=(-c/n)D_1^2 \beta_2$ dominates the overall cavity dispersion. Some parameters such as the dispersion and mode volume are obtained with the finite-element simulation~\cite{Fujii2020}. To take account of thermal effects in the simulation, the detuning term can be rewritten as $\delta_\mathrm{eff}$, which includes the detuning shift induced by the thermo-optic (TO) effect~\cite{Li2017}. It should be noted that no distinct influence is observed owing to the low thermal conductivity and the relatively large size of the AMTIR-1 resonator. The details of LLE simulations including thermal effects are provided in Supplement~1. Figure~\ref{Fig_dispersion}(b) shows the optical spectrum and integrated dispersion $D_\mathrm{int}/2\pi$ when an AMTIR-1 resonator with an FSR of 20~GHz is driven with a pump power of 500~mW. The weak anomalous dispersion profile and dispersive wave emission allows a broadband comb spectrum across $6-8$~\textmu m wavelength. The evolution of the intracavity power is shown in Fig.~\ref{Fig_dispersion}(c), as a function of the normalized detuning. These results show that a single soliton state appears after discrete power steps, indicating the transitions to stable DKS states from chaotic MI combs. It is worth mentioning that the loaded Q-factor is only $7.5\times10^6$ thanks to the significantly high nonlinear coefficient, which is three orders of higher than silica or $\mathrm{MgF_2}$, both of which are widely used cavity materials for DKS generation.

In conclusion, we have demonstrated monolithic high-Q WGM resonators made of AMTIR-1 chalcogenide glass. The observed loaded Q exceeds $10^7$ in the 1550~nm band, and to the best of our knowledge this is the highest reported value for AMTIR-1 microcavities. We revealed that the large discrepancy between the thermal resonance tuning rates of the experiment and theoretical prediction can mainly be attributed to the intrinsic low thermal conductivity of AMTIR-1. Finally, we numerically simulated DKS microcomb generation at a wavelength of 7~\textmu m, which balances a moderate anomalous dispersion and a high-Q property. For future experimental efforts, quantum cascade lasers would be promising candidates for the pump source~\cite{Schwarz2017}, owing to their high output power and broad tunability in the mid-infrared region. High transparency as well as the high optical nonlinearity of AMTIR-1 make this resonator a highly attractive platform for mid-infrared and long-wave infrared nonlinear optics including quantum state generation and miniature frequency comb generation, which are indispensable for chemical sensing and molecular spectroscopy.

\vskip\baselineskip
See Supplement 1 for supporting content.


\section*{Acknowledgments}
This work was supported by JSPS KAKENHI (JP24K17624); Adaptable and Seamless Technology transfer Program through Target-driven R\&D (A-STEP) from Japan Science and Technology Agency (JST) (JPMJTR23RF); Keio University Program for the Advancement of Next Generation Research Projects; Inamori Foundation; L. Yang acknowledges China Scholarship Council (202306280149). R. Sugano acknowledges JST SPRING (JPMJSP2123). We are grateful for the technical support provided by the Manufacturing Center at Keio University.

\section*{Disclosures}
The authors declare no conflicts of interest.

\section*{Data availability} Data underlying the results presented in this paper are obtained from the authors upon reasonable request.


\bibliography{amtir}


\end{document}